\documentclass[doublespacing]{elsart}

\usepackage{amssymb,epsfig}

\newcommand{\cH}{\mathcal{H}}
\newcommand{\cM}{\mathcal{M}}

\newcommand{\br}{{\bf r}}

\newcommand{\tr}{\,{\rm tr\/}\,}
\newcommand{\lp}{\left(}
\newcommand{\rp}{\right)}
\newcommand{\lb}{\left[}
\newcommand{\rb}{\right]}
\newcommand{\lc}{\left\{}

\newcommand{\lbb}{\langle\langle}
\newcommand{\rbb}{\rangle\rangle}
\newcommand{\eq}[1]{Eq.\ (\ref{#1})}

\begin{document}

\begin{frontmatter}

\title{Magnetization and spin-spin energy diffusion in the XY model: a diagrammatic approach}

\author{Daniel Greenbaum}

\ead{daniel.greenbaum@weizmann.ac.il}

\address{Department of Physics, Massachusetts Institute of
Technology, Cambridge, Massachusetts 02139\\Department of
Condensed Matter Physics, Weizmann Institute of Science, 76100
Rehovot, Israel\thanksref{label1}}

\thanks[label1]{present address}

\begin{abstract}
It is shown that the diagrammatic cluster expansion technique for
equilibrium averages of spin operators may be straightforwardly
extended to the calculation of time-dependent correlation
functions of spin operators. We use this technique to calculate
exactly the first two non-vanishing moments of the spin-spin and
energy-energy correlation functions of the XY model with arbitrary
couplings, in the long-wavelength, infinite temperature limit
appropriate for spin diffusion. These moments are then used to
estimate the magnetization and spin-spin energy diffusion
coefficients of the model using a phenomenological theory of
Redfield. Qualitative agreement is obtained with recent
experiments measuring diffusion of dipolar energy in calcium
fluoride.
\end{abstract}

\end{frontmatter}


\section{Introduction}

Experimentally measured quantities in spin systems can often be
expressed in terms of time-dependent correlation functions of spin
operators.\cite{Abragam,AG} A well-known example is the free
induction decay lineshape in solids.\cite{VanVleck,LN} Another
example is the rate of spin diffusion,\cite{Bloembergen} the
transport of magnetization or spin-spin energy by mutual flips of
spin pairs having the same Zeeman splitting.

The calculation of time-dependent correlation functions can be
challenging both because of the structure of typical Hamiltonians
for spin-spin interactions and because of the non-trivial
commutation properties of spin operators. Of particular difficulty
is the analysis of correlation functions of more than two spin
operators. These arise in studying the diffusion of spin-spin
energy, a problem in which interest has been revived by recent
experiments that directly observed the diffusion of magnetization
and dipolar energy in calcium fluoride.\cite{Zhang,Boutis} A
phenomenological approach developed by de Gennes\cite{dG} and
Redfield\cite{Redfield} to calculate spin diffusion coefficients
based on the knowledge of the first few moments of the associated
correlation functions agrees well with experiments on
magnetization diffusion. However, because of the difficulty of
calculating moments for systems with long-range interactions, such
as calcium fluoride, this approach has not been used to study
spin-spin energy diffusion in such systems, while magnetization
diffusion has only been studied to lowest order in perturbation
theory in the flip-flop (or XY) term of the Hamiltonian.\cite{RY}

In this paper we present a diagrammatic technique for calculating
the moments of time-dependent correlation functions, allowing a
simplified treatment of the type of problems mentioned above. This
technique extends an approach originally developed for calculating
static, equilibrium averages in spin
systems.\cite{Brout1,Brout2,Englert,SHEB} The extension is based
on the cancellation of disconnected diagrams, proved in Appendix\
\ref{appb}. This cancellation greatly reduces the number of
diagrams one needs to consider, and constitutes the primary
advantage of the method.

The method is illustrated through application to the XY model.
This model was chosen because it contains the simplest Hamiltonian
exhibiting the dynamics of spin diffusion -- mutual flips of spin
pairs. It is therefore expected to qualitatively reproduce the
behavior of more complicated systems, such as dipolar-coupled
spins in high field, for which the dynamics is governed by the
spin-flip process. It is also useful for comparison to the
perturbative limit.\cite{RY} We calculate the first two
non-vanishing moments of the spin-spin and energy-energy
correlation functions in this model for arbitrary couplings, at
infinite temperature. The expressions are exact in the
long-wavelength limit. From these moments, analytic expressions
for the diffusion coefficients are obtained. Choosing the coupling
constants in our calculation to be those of calcium fluoride gives
numerical values in qualitative agreement with experiments, as
shown in Table\ \ref{moments-results}. The ratio we find for the
diffusion coefficients of magnetization and spin-spin energy is,
however, a few times smaller than experimentally
measured.\cite{Boutis} This may be due to our not having
considered the full dipolar interaction, or to the importance of
coherences in the quantum state of the spin system,\cite{DG} which
would not be taken into account by the present approach.

Besides the moment method, other approaches have yielded spin-spin
energy diffusion coefficients, such as non-equilibrium statistical
mechanics\cite{BW1} and classical simulations.\cite{TW} However,
the assumptions and approximations involved were difficult to
justify and gave results which were not in better agreement with
the recent experiments than those found here. Another recent
calculation\cite{DG} for dipolar interactions was limited to the
first two orders of perturbation theory in the flip-flop (XY)
interaction, and gave similar qualitative agreement with the
experiments. The work presented here should therefore complement
the previous studies.

\section{Model}

The XY-model for $N$ spins on a rigid lattice is
\begin{equation}
\cH = \sum_{i,j}^N B_{ij}I_i^+I_j^-, \label{hamiltonian}
\end{equation}
with $B_{ii}=0$ (no sum). The latin indices run over all lattice
sites and the $I_i^\alpha$ are spin operators defined by their
commutation relations $[I_i^\alpha,I_j^\beta] =
\delta_{ij}I_i^\gamma$, where $\alpha,\,\beta,\,\gamma$ is any
cyclic permutation of x, y, z. The $I_j^\pm \equiv I_j^x \pm
iI_j^y$ are raising and lowering operators. The combination of
operators, $I_i^+I_j^-$, generates mutual flips of spin pairs
which are responsible for the transport of magnetization and
spin-spin energy (or heat). The coefficients $B_{ij}$ ($i\neq j$)
are arbitrary. To make contact with dipolar coupled spins, we use
\begin{equation}
B^{dip}_{ij} = \frac{\gamma^2 \hbar}{4} \frac{3
\cos^2\theta_{ij}-1}{r_{ij}^3}. \label{Bdipij}
\end{equation}
Here $\gamma$ is the nuclear gyromagnetic ratio, $\br_{ij}$ is the
displacement between lattice sites $i$ and $j$, and $\theta_{ij}$
is the angle between $\br_{ij}$ and the external magnetic field
${\bf B}_0$, which is taken to lie along the $z$-axis. We do not
include the Zeeman energy in \eq{hamiltonian} as it may be
eliminated by a unitary transformation to the rotating
frame.\cite{Abragam}

The full Hamiltonian for dipolar coupled spins in a strong
magnetic field\cite{Abragam} contains an additional term
$-2\sum_{i,j}^N B^{dip}_{ij}I_i^zI_j^z$, which we ignore here as
discussed above. A complementary approach which includes this term
but is perturbative in \eq{hamiltonian} has been discussed
earlier.\cite{DG}

The quantities of physical interest are correlation functions of
the form
\begin{equation}
c_{S}(k,t) = \frac{\langle S(-k,t)S(k,0)\rangle}{\langle
S(-k,0)S(k,0)\rangle}, \label{correlator}
\end{equation}
where $S(k,t) = \sum_i e^{ikz_i} S_i(t)$ is a spin operator or
product of spin operators in the Heisenberg representation, and
$k$ is the magnitude of the wavevector, which points along the
magnetic field axis. In the specific cases which we consider
below, $S_i$ is either the local magnetization, $S_i =
-\gamma\hbar I_i$, or spin-spin energy, $S_i = \sum_{j, (j\neq i)}
\cH_{ij}$, at lattice site $i$. The angular brackets denote
averaging over an equilibrium ensemble. For most NMR problems,
including spin diffusion, it suffices to consider $T=\infty$, so
that $\langle \cdots \rangle = \tr\{\cdots\}/\tr\{{\bf 1}\}$. The
extension to finite temperature is straightforward and will not be
considered here.

Below we will be interested in the moments of the correlation
function, \eq{correlator}. Expanding in Taylor series about $t =
0$, we obtain
\begin{equation}
c_{S}(t) = \sum_{n = 0}^{\infty} \frac{1}{(2n)!} M_S^{(2n)}
t^{2n}. \label{expansion}
\end{equation}
The even moments $M_S^{(2n)}$ are given by
\begin{equation}
M_S^{(2n)} = (-1)^n \left( \frac{1}{\hbar} \right)^{2n}
\frac{\langle S(k,0)[\cH,S(-k,0)]_{2n}\rangle}{\langle
S(-k,0)S(k,0)\rangle }, \label{moment}
\end{equation}
where $[A,B]_{n} \equiv [A,[A,[...[A,B]...]]]$, with $A$ appearing
$n$ times. The sum in \eq{expansion} involves only even powers of
$t$ because the odd moments are zero. These expressions may be
derived by expanding $S(k,t) = e^{i\cH t}S(k,0)e^{-i\cH t}$ by the
well-known formula $e^{A}Be^{-A} = \sum_{n=0}^{\infty}
\frac{1}{n!} [A,B]_{n}$ and putting the result in \eq{correlator}.

Following Redfield,\cite{Redfield} one can obtain an approximate
value of the diffusion coefficient of $S$ from the first two
non-vanishing moments,\cite{Lorentzian}
\begin{equation}
D_S = \frac{1}{k^2\tau_S} = \frac{1}{k^2}
\sqrt{\frac{\alpha_4}{\alpha_2^3}\frac{(M^{(2)})^3}{M^{(4)}}},\label{D}
\end{equation}
where $\alpha_2$ and $\alpha_4$ are certain phenomenological
parameters. We show in the next section that each moment is
proportional to $k^2$ at long wavelength, so that this expression
for $D_S$ is independent of $k$. \eq{D} is obtained by matching
the terms of \eq{expansion} to a phenomenological decay function
of the form $f_S(t) = g_S(t)e^{-t/\tau_S}$, where $\tau_S =
(k^2D_S)^{-1}$ is the diffusion time. The cutoff function $g_S(t)$
is different from unity only at times short compared to the
spin-spin correlation time, $T_S \equiv \hbar/{\rm max}(B_{ij})$,
which is roughly the time required for a single spin flip, and is
in principle determined by the microscopic dynamics.

The exact values of the parameters $\alpha_{2n}$ are related to
the manner in which the cutoff function $g_S(t)$ vanishes at high
frequency. For example, a Gaussian and step-function cutoff give
\begin{equation}
\alpha_{2n} = \lc \begin{array}{cc}
\frac{(-1)^n(2n-2)!}{\sqrt{\pi}2^{2n-2}(n-1)!}, & g_S(\omega) =
e^{-\omega^2T_S^2}, \\
\frac{(-1)^n2}{\pi(2n-1)}, & g_S(\omega) = \Theta(T_S^{-1} -
\omega).\end{array}\right. \label{alphas}
\end{equation}
Both values have the same order of magnitude. Here $g_S(\omega)$
is the Fourier transform of $g_S(t)$. Since the shape of the
cutoff function is not determined within the phenomenological
model, \eq{D} can only be viewed as approximate. Nevertheless,
this shape is not expected to be drastically different for the
magnetization and spin-spin energy diffusion coefficients, and
therefore their ratio can be expected to have a weaker dependence
on cutoff.

\section{Calculation of moments}
In this section we calculate the second and fourth moments of
magnetization and spin-spin energy for the XY model. Since we are
interested in the long-wavelength behavior, we Taylor expand the
correlation function, \eq{correlator}, in $k$. This gives
\begin{eqnarray}
c_S(k,t) &=& \frac{\sum_{i,j}e^{ik(z_i-z_j)}\langle
S_i(0)S_j(t)\rangle }{\sum_i\langle S_i(0)^2\rangle }
\nonumber \\
&\simeq & 1 - \frac{k^2}{2}\frac{\sum_{i,j}z_{ij}^2\langle
S_i(0)S_j(t)\rangle }{\sum_i\langle S_i(0)^2\rangle } + O(k^4),
\end{eqnarray}
where $z_{ij} \equiv z_i - z_j$, and the terms odd in $z_{ij}$ are
zero. The $O(k^4)$ term is safely neglected as the correlation
$\langle S_i(0)S_j(t)\rangle $ is a rapidly decaying function of
the distance $|\br_i - \br_j|$. It depends on products of the
spin-spin couplings, $B_{ij}$, which are either short-ranged or,
in the case of dipolar coupling, decay algebraically on a length
scale of a few lattice spacings. We will demonstrate this
explicitly for each moment. The wavelength, $\lambda = 2\pi/k$, is
taken to be much longer than this decay scale. In the calcium
fluoride experiments\cite{Zhang,Boutis} it is at least $10^4$
lattice spacings. Expanding the commutator in \eq{moment}, we
obtain
\begin{eqnarray}
M_S^{(2n)} = \frac{(-1)^{n+1}k^2}{2\sum_i\langle S_i(0)^2\rangle
}\sum_{i,j}z_{ij}^2 \sum_{m=0}^{2n}\lp
\begin{array}{c} 2n \\ m
\end{array}\rp (-1)^m \langle \cH^m S_j(0) \cH^{2n-m}S_i(0)\rangle ,
\label{moment-binomial}
\end{eqnarray}
for $n\ge 1$. Here $\lp \begin{array}{c} 2n \\ m
\end{array}\rp = \frac{(2n)!}{m!(2n-m)!}$ is a binomial
coefficient, and we have used
\begin{equation}
[\cH,S_j(0)]_{2n} = \sum_{m=0}^{2n}\lp \begin{array}{c} 2n \\ m
\end{array}\rp (-1)^m \cH^m S_j(0) \cH^{2n-m}.
\end{equation}
\eq{moment-binomial} proves the $k^2$ dependence mentioned in the
last section.

To calculate the moments for the XY model from
\eq{moment-binomial}, one must evaluate averages of the form
$\langle \cH^m S_j(0) \cH^{2n-m}S_i(0)\rangle $. We do this using
a diagrammatic cluster-expansion
technique,\cite{Brout1,Brout2,Englert,SHEB,AG} extended to
\eq{moment-binomial} with the help of a theorem proved in
Appendix\ \ref{appb}. This technique eliminates the need for
keeping track of the Kronecker deltas that arise from the
contractions of spin operators, and allows the identification of
the most important contributions to \eq{moment-binomial} at each
$n$. It is based on an ordered cumulant expansion of spin operator
averages. For completeness, a brief introduction to ordered
cumulants of spin operators, also known as
semi-invariants,\cite{Brout1,Brout2,Englert,SHEB,AG} is given in
Appendix\ \ref{cumulant-app}.

\subsection{Magnetization moments}
\label{magnetization-moments}

Let $S_i = -\gamma\hbar I_i^z$, and consider the expression,
\begin{equation}
T_{2n} = (-1)^{n+1}\sum_{i,j}z_{ij}^2\sum_{m=0}^{2n}\lp
\begin{array}{c} 2n \\ m
\end{array}\rp (-1)^m \langle \cH^m I_j^z \cH^{2n-m}I_i^z\rangle ,
\label{trace-term}
\end{equation}
in the numerator of \eq{moment-binomial}. A diagram element is
associated to each operator in this expression as follows.
\begin{eqnarray}
I_i^z \; \longrightarrow \; \epsfig{file=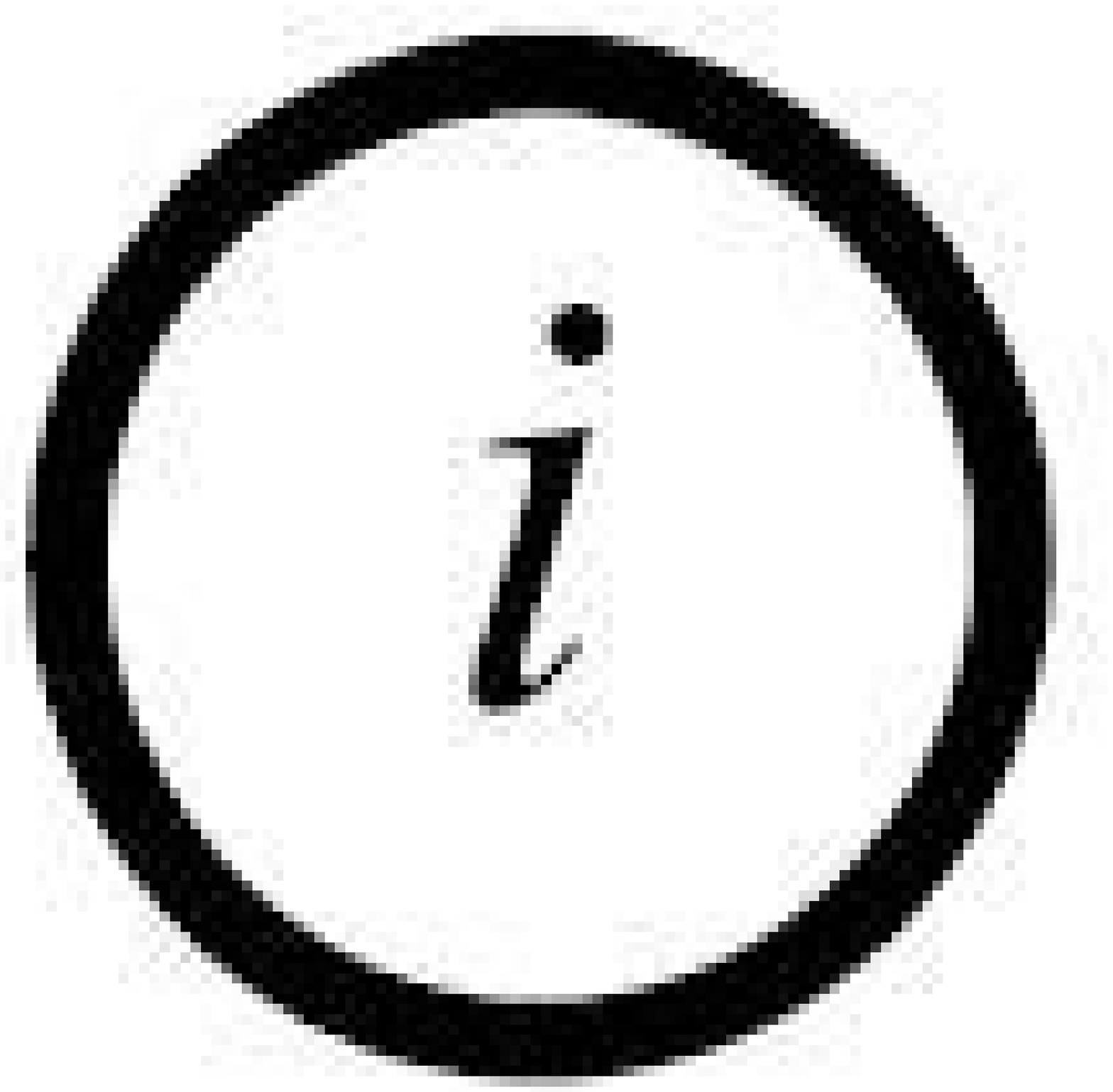,height=2ex}
\label{Iz-diagram}\\
\cH = \sum_{kl} B_{kl} I_k^+ I_l^- \; \longrightarrow \;
\epsfig{file=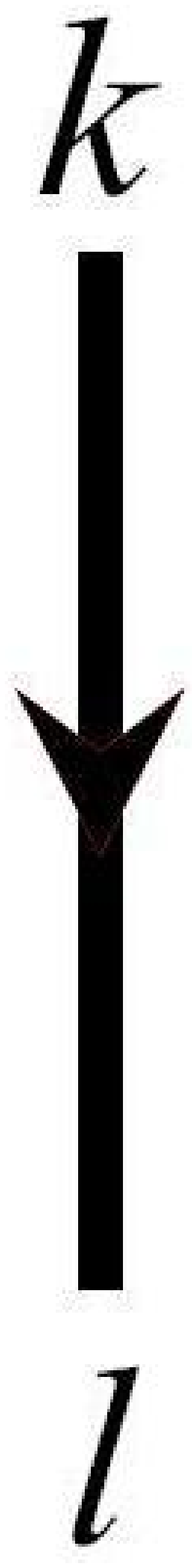,height=10ex} \label{h-diagram}
\end{eqnarray}
The indices $k,\ l$ are dummies that are summed over, and in
practice can be left off of diagrams. One then forms all possible
topologically distinct, connected diagrams from these elements by
joining them end-to-end in all possible ways, with the open
circles for $I^z$ inserted at vertices. The diagram elements are
numbered based on the order in which they appear in
\eq{trace-term}. This order must be kept track of because of the
non-trivial commutation properties of spin operators. For example,
the diagrams corresponding to $\langle I_i^z\cH I_j^z
\cH\cH\cH\rangle $ are numbered as follows.
\begin{center}
\epsfig{file=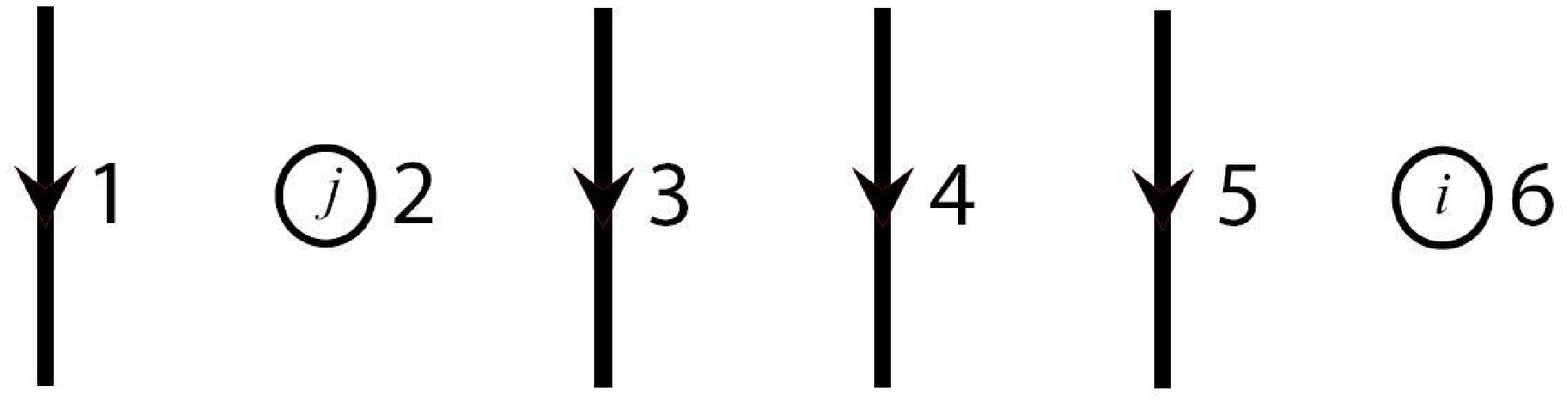,height=8ex}
\end{center}

Each vertex without a circle is assigned a dummy summation index,
and each vertex with a circle receives the index corresponding to
that circle. The circles corresponding to $i$ and $j$ must appear
at different vertices, since the $z_{ij}^2$ factor in
\eq{trace-term} ensures that $i\neq j$. To each vertex is assigned
an ordered cumulant. Each interaction line has an interaction
coefficient associated with it that has the appropriate indices.
E.g. the line $\ \epsfig{file=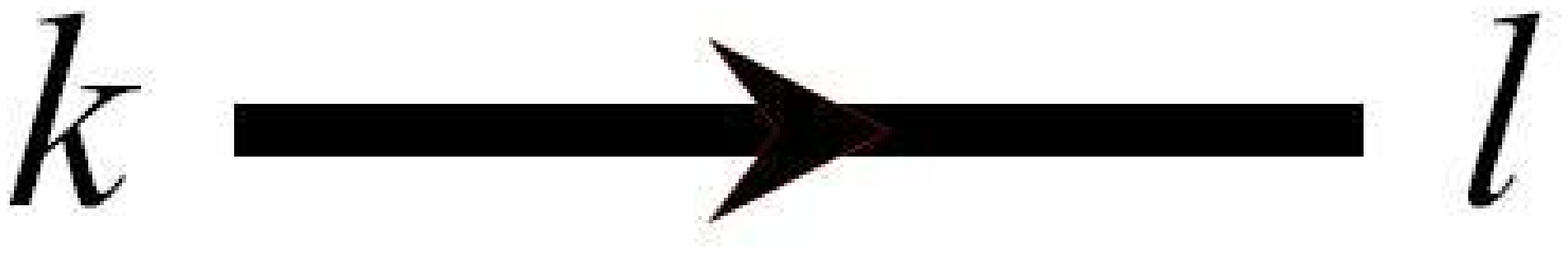,width=8ex}\ $
receives a factor of $B_{kl}$. The analytic expression
corresponding to a given diagram is formed by taking the product
of all the ordered cumulants and interaction coefficients
associated with it, and summing over all dummy indices without
restriction. The sum includes a factor of $z_{ij}^2$ and the
appropriate binomial coefficients appearing in \eq{trace-term}.

Of the total set of possible diagrams, many do not contribute.
There are no diagrams with free ends, as these represent
uncontracted spin operators which cause the trace to vanish. Each
vertex must have the same number of lines leaving as entering,
since all ordered cumulants with an unequal number of raising and
lowering operators are zero. Finally, the disconnected diagrams
vanish, as shown in Appendix \ref{appb}.

\begin{figure}
\begin{center}
\epsfig{file=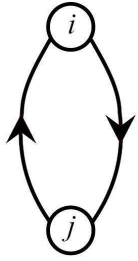,width=10ex}
\end{center}
\caption{Diagram contributing to second moment for magnetization}
\label{MZ2-diagram}
\end{figure}

The only diagram contributing to the second moment is shown in
Fig. \ref{MZ2-diagram}. Its contribution to \eq{trace-term} is
\begin{eqnarray}
T_2 &=& \sum_{ij}z_{ij}^2\sum_{m=0}^2 \lp
\begin{array}{c} 2 \\ m
\end{array}\rp (-1)^m \times \epsfig{file=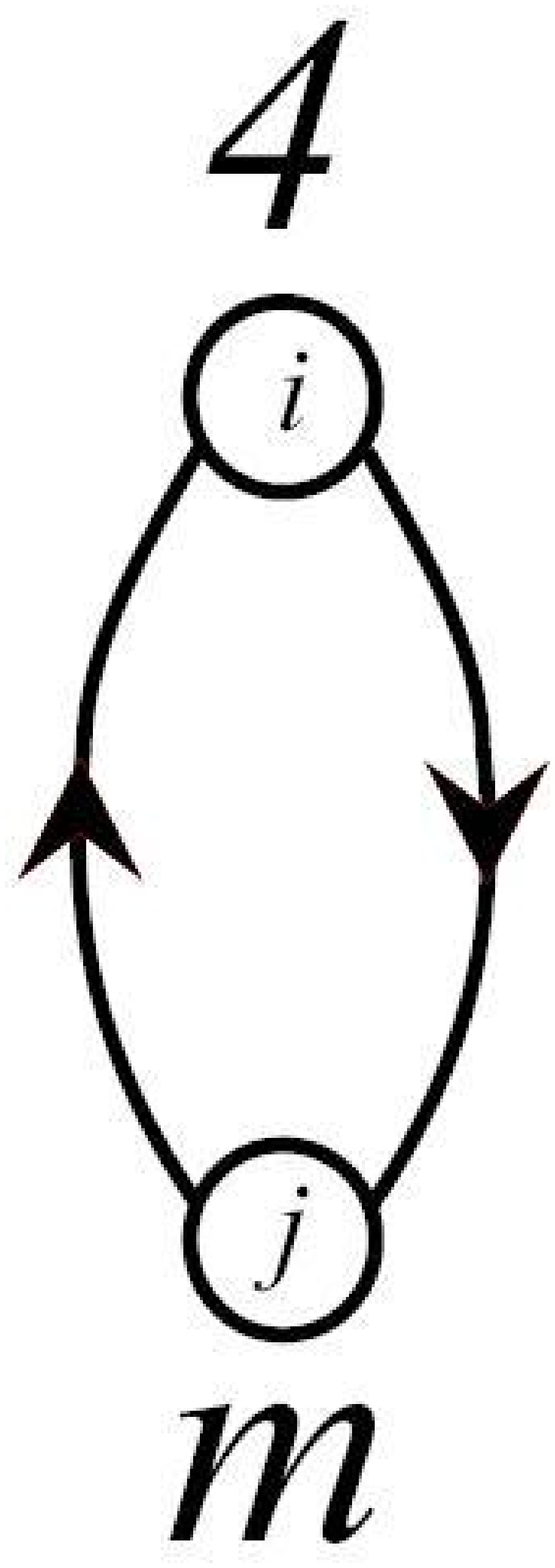,height=10ex}
\nonumber \\
&=&\hspace{-1ex}\sum_{ij}z_{ij}^2 \times
\epsfig{file=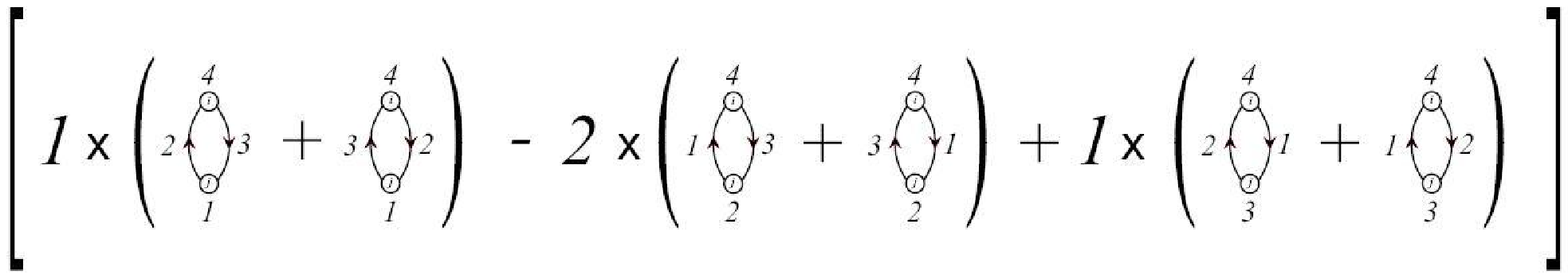,height=10ex}
\nonumber \\
&=&\hspace{-1ex}\sum_{ij}z_{ij}^2B_{ij}^2 \lb\lp\vphantom{\int}
\lbb z+- \rbb \lbb -+z \rbb + \lbb z-+ \rbb \lbb +-z \rbb \rp
\right.
\nonumber \\
&& - 2 \lp \vphantom{\int} \lbb +z- \rbb \lbb
-+z \rbb + \lbb -z+ \rbb \lbb +-z \rbb \rp \nonumber \\
&&  + \left.\lp\vphantom{\int} \lbb -+z \rbb \lbb
+-z \rbb + \lbb +-z \rbb \lbb -+z \rbb \rp\rb \nonumber \\
&=& -\frac{1}{2}\sum_{ij}z_{ij}^2B_{ij}^2.
\end{eqnarray}
The values of the ordered cumulants are $\lbb +-z \rbb =
\frac{1}{4} $ and $\lbb -+z \rbb = -\frac{1}{4}$, as given in
Table \ref{cumulants}.

The denominator of \eq{moment-binomial} may be calculated without
diagrams, and we obtain
\begin{equation}
\sum_i \langle (I_i^z)^2\rangle  = \frac{N}{4}.
\end{equation}
Inserting these results into \eq{moment-binomial} gives
\begin{equation}
M_{\cM}^{(2)} = -k^2\sum_i z_{ik}^2B_{ik}^2. \label{MZ2}
\end{equation}
We note that, because of translational invariance, we can drop the
summation over the dummy index $k$.

\begin{figure}
\begin{center}
\epsfig{file=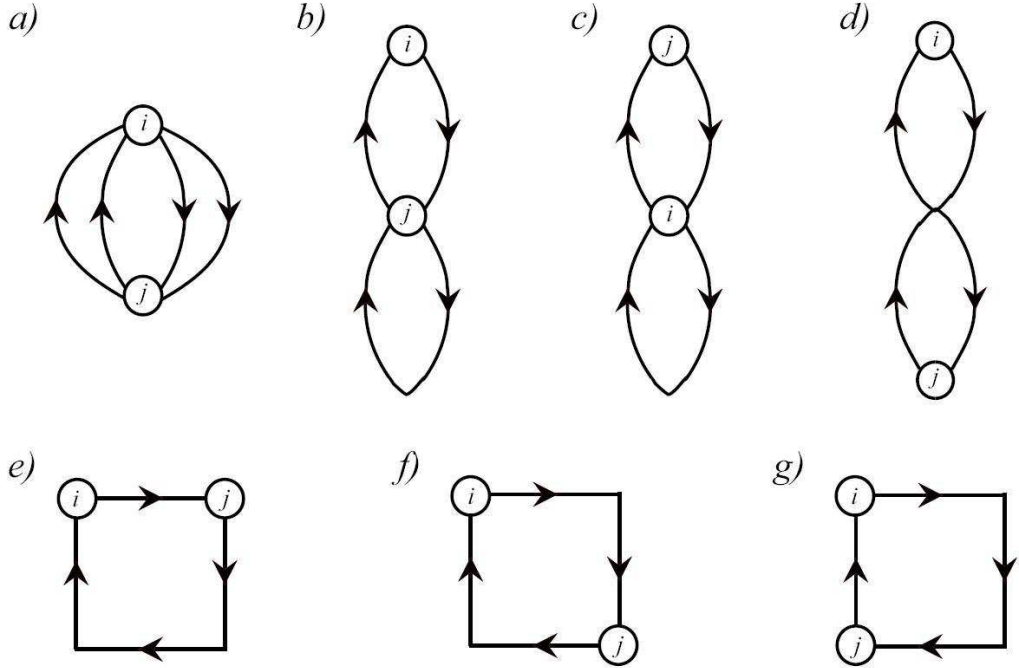,height=50ex}
\end{center}
\caption{All topologically distinct diagrams containing two
circles and four interaction lines. The diagrams shown here arise
in the calculation of the fourth moment for magnetization as well
as that of the second moment for spin-spin energy. The analytic
expressions for the diagrams are different in the two cases,
however.} \label{MZ4-diagrams}
\end{figure}

The diagrams contributing to the fourth moment for magnetization
are shown in Fig. \ref{MZ4-diagrams}. They are calculated in a
similar way to those for the second moment, so we omit the
details. Table \ref{cumulants} shows that most of the fourth and
fifth-order cumulants are zero, which enables us to consider only
a subset of the orderings of the diagram elements. The non-zero
cumulants at fourth and fifth order correspond to vertices with
two ingoing and two outgoing lines, with both ingoing lines next
to each other in the order (same for the outgoing lines). The
calculation shows that only the diagrams labelled $a)$, $b)$, and
$c)$ in Fig. \ref{MZ4-diagrams} contribute. The fourth moment is
\begin{equation}
M_{\cM}^{(4)} = -4k^2\lb \sum_i z_{ik}^2 B_{ik}^4 - \lp \sum_i
z_{ik}^2 B_{ik}^2 \rp \lp \sum_i B_{ik}^2 \rp \rb. \label{MZ4}
\end{equation}

\subsection{Energy moments}
\label{energy-moments}

The expression in the numerator of \eq{moment-binomial} for
spin-spin energy, corresponding to \eq{trace-term}, is
\begin{eqnarray}
T_{2n} = (-1)^{n+1}\sum_{i,j,k,l}z_{ij}^2\sum_{m=0}^{2n}\lp
\begin{array}{c} 2n \\ m
\end{array}\rp (-1)^m \langle \cH^m \cH_{jl}
\cH^{2n-m}\cH_{ik}\rangle, \label{trace-term-energy}
\end{eqnarray}
where $S_i = \sum_{k, (k\neq i)} \cH_{ik}$, and
\begin{eqnarray}
\cH_{ik} &=& \cH_{ik}^{(+)} + \cH_{ik}^{(-)}, \label{hplusminus} \\
\cH_{ik}^{(+)} &\equiv & \frac{1}{2}B_{ik} I_i^+ I_k^-, \label{hplus}\\
\cH_{ik}^{(-)} &\equiv & \frac{1}{2}B_{ik} I_i^- I_k^+.
\label{hminus}
\end{eqnarray}
We can rewrite \eq{trace-term-energy} as
\begin{eqnarray}
T_{2n} &=& 2\lp T_{2n}^{(+)} + T_{2n}^{(-)}\rp \label{trace-term+-} \\
T_{2n}^{(+)} &\equiv & \sum_{i,j,k,l}z_{ij}^2\sum_{m=0}^{2n}\lp
\begin{array}{c} 2n \\ m
\end{array}\rp (-1)^{(n+m+1)} \langle \cH^m \cH_{jl}^{(+)}
\cH^{2n-m}\cH_{ik}^{(+)}\rangle , \label{trace-term+}\\
T_{2n}^{(-)} &\equiv & \sum_{i,j,k,l}z_{ij}^2\sum_{m=0}^{2n}\lp
\begin{array}{c} 2n \\ m
\end{array}\rp  (-1)^{(n+m+1)} \langle \cH^m \cH_{jl}^{(+)}
\cH^{2n-m}\cH_{ik}^{(-)}\rangle , \label{trace-term-}
\end{eqnarray}
where use has been made of the formula
\begin{equation}
\langle (A + A^\dag)(B + B^\dag)\rangle  = 2 {\rm Re}\ \langle (A
+ A^\dag)B\rangle ,
\end{equation}
for any operators $A$ and $B$.

We associate the following diagram elements with the operators
appearing in Eqs.\ (\ref{trace-term+}) and (\ref{trace-term-}).
\begin{eqnarray}
\cH_{ik}^{(+)} \; \longrightarrow \;
\epsfig{file=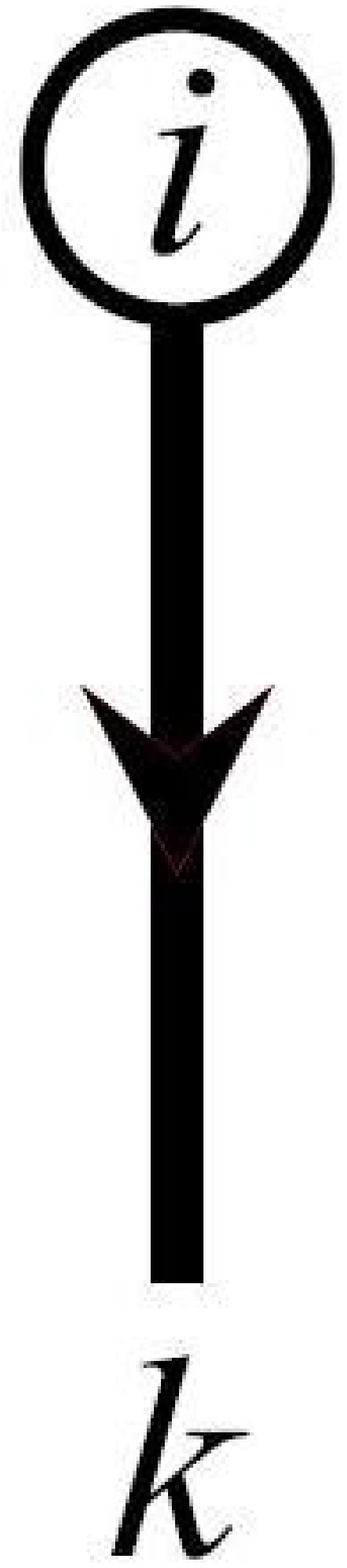,height=10ex} \label{hplus-diagram}
\\ \nonumber \\
\cH_{ik}^{(-)} \; \longrightarrow \;
\epsfig{file=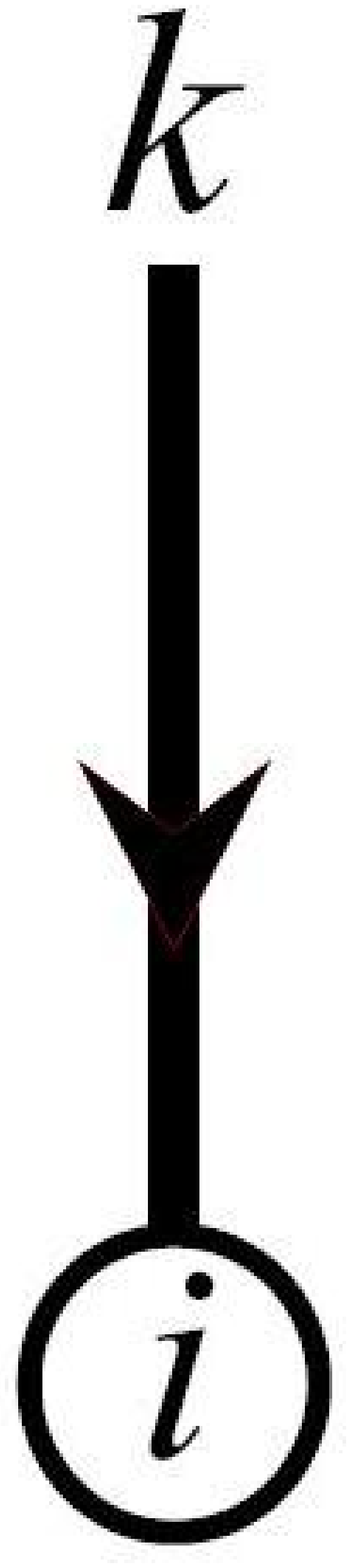,height=10ex} \label{hminus-diagram}
\end{eqnarray}
The diagram element for the full interaction, $\cH$, is the same
as in the last section, i.e. \eq{h-diagram}. Dummy indices such as
$k$ will be left off of the diagrams as before.

The calculation of \eq{trace-term-energy} is similar to that of
\eq{trace-term}. In this case, however, the interaction lines due
to $\cH_{ik}^{(+)}$ and $\cH_{ik}^{(-)}$ receive an additional
factor of $\frac{1}{2}$, because this factor appears in Eqs.\
(\ref{hplus}) and (\ref{hminus}). The final result is multiplied
by the factor $2$ appearing in \eq{trace-term+-}.

The diagrams contributing to the second moment for spin-spin
energy are shown in Fig. \ref{MZ4-diagrams}. These diagrams are
exactly the same as the ones arising in the calculation of the
fourth moment for magnetization. However, their meaning is
different, as now there are no $I^z$ operators, and we deal with a
different set of ordered cumulants. We note that the diagrams at
order (2n) for spin-spin energy are always the same as those at
order (2n+2) for magnetization.

\begin{figure}
\begin{center}
\epsfig{file=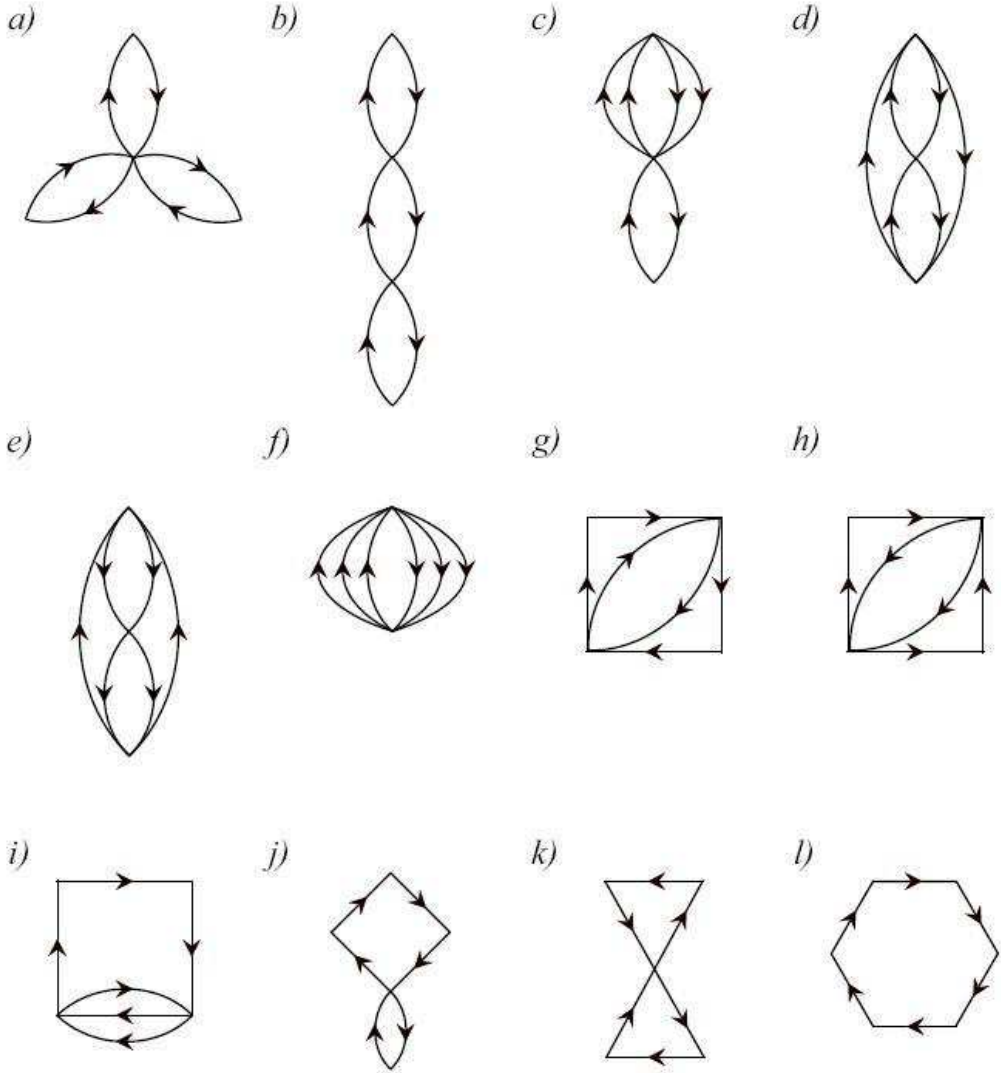,width=75ex}
\end{center}
\caption{All topologically distinct diagrams containing six
interaction lines. Diagrams for the fourth energy moment are
obtained by placing circles with indices $i$ and $j$ at vertices
in all distinct ways.} \label{MH4-diagrams}
\end{figure}

One can easily see that the diagrams labelled $e)$, $f)$, and $g)$
in Fig. \ref{MZ4-diagrams} are zero. Associated with each of them
is the product of ordered cumulants, $\lbb +- \rbb^4 =
\frac{1}{16}$. Because this cumulant factor is the same regardless
of the order of diagram elements, we can move all the diagrams to
the left of the second summation sign in \eq{trace-term-energy}.
For example, diagram $e)$ gives
\begin{eqnarray}
T_2(e) &=&\hspace{-1ex}2\sum_{ij}z_{ij}^2\sum_{m=0}^2 \lp
\begin{array}{c} 2 \\ m
\end{array}\rp (-1)^m \times
\epsfig{file=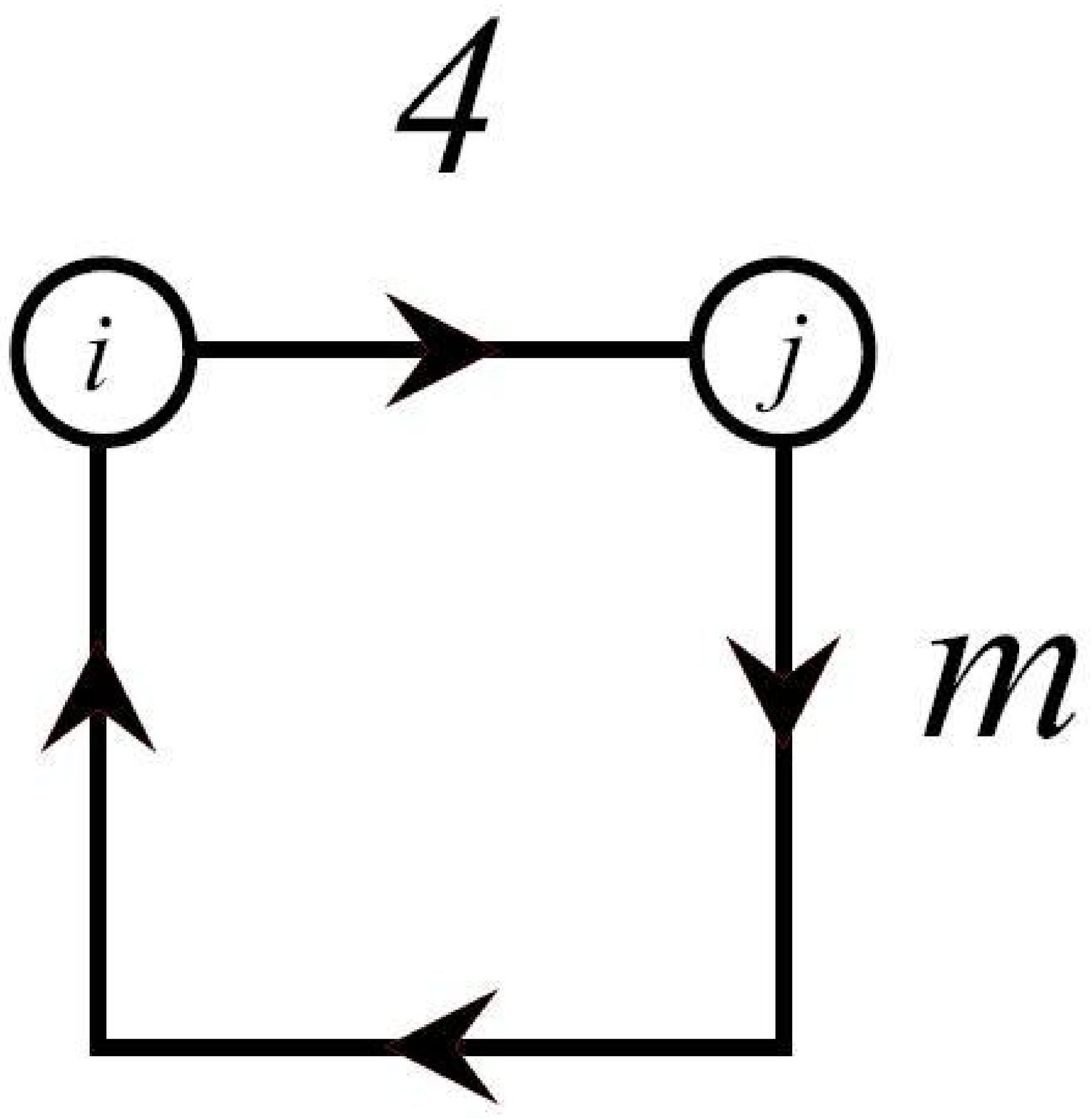,height=10ex} \nonumber \\
&=&\hspace{-1ex}2\hspace{-1ex}\sum_{ij}z_{ij}^2
\times\hspace{-1ex}
\epsfig{file=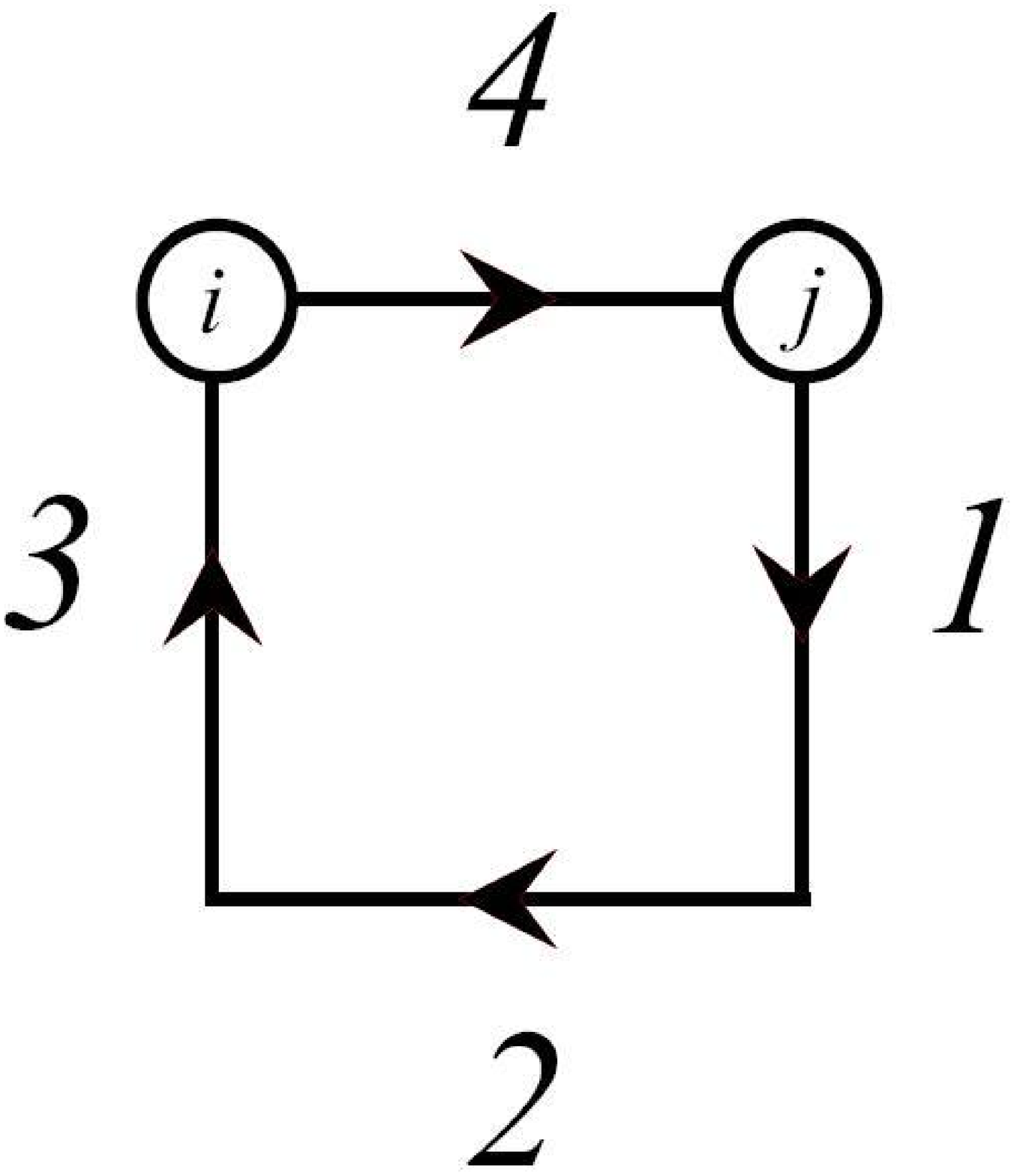,height=12ex}\hspace{-1ex}
\times\hspace{-1ex}\sum_{m=0}^2 \lp
\begin{array}{c} 2 \\ m
\end{array}\rp (-1)^m.
\end{eqnarray}
Since the sum over binomial coefficients is zero, we have $T_2(e)
= 0$.

By direct calculation, it is also easily found that the diagrams
labelled $a)$, $b)$, and $c)$ are zero. The only diagram
contributing to the second moment for spin-spin energy is
therefore diagram $d)$ of Fig. \ref{MZ4-diagrams}.
\eq{trace-term-energy} then reads
\begin{equation}
T_2 = 2\sum_{ij}z_{ij}^2\sum_{m=0}^2 \lp
\begin{array}{c} 2 \\ m
\end{array}\rp (-1)^m \times
\epsfig{file=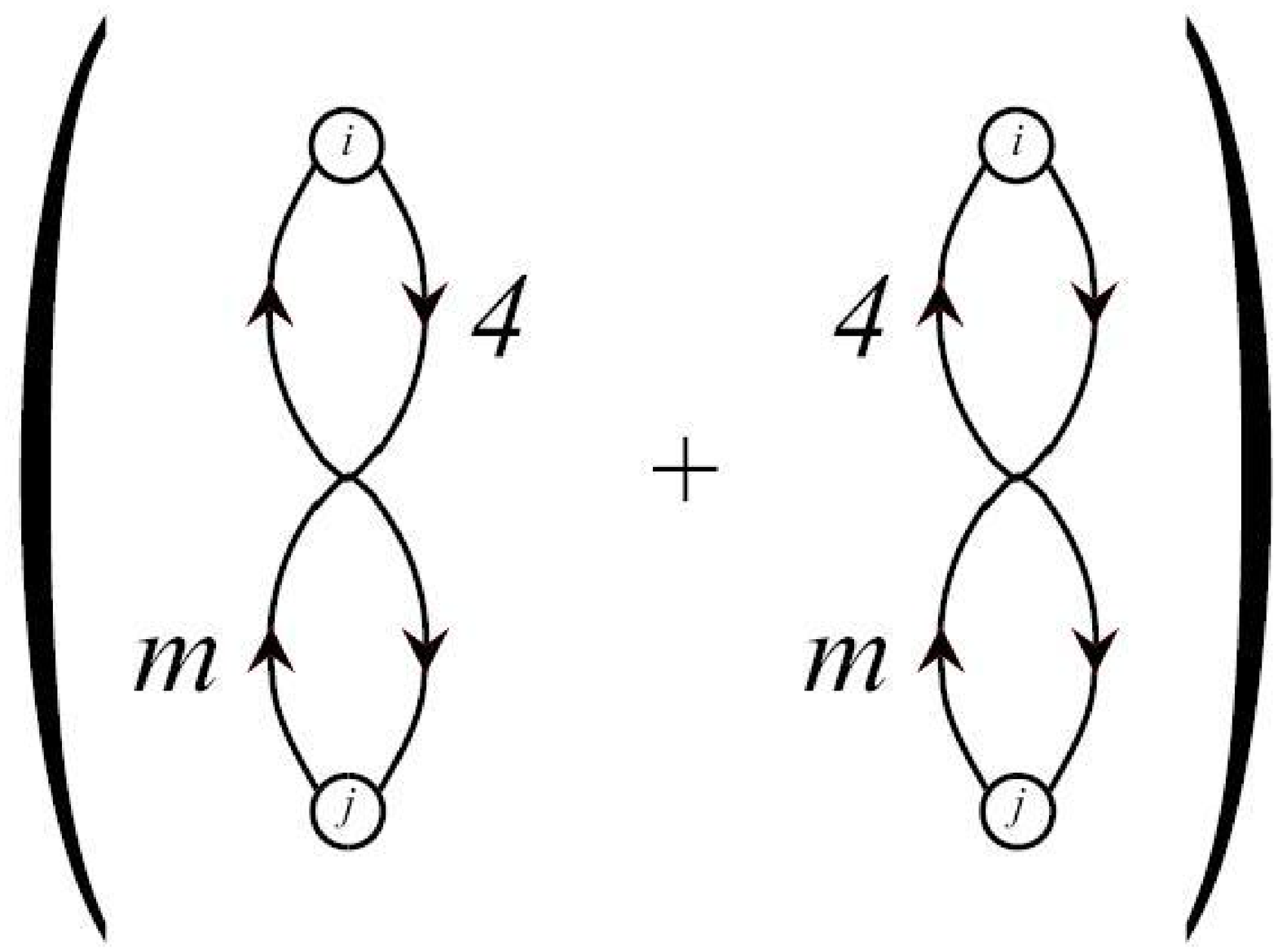,height=10ex}.
\label{trace-term-d}
\end{equation}
According to Table \ref{cumulants}, $\lbb +-+- \rbb = 0$. This
restricts the possible orderings of the diagram elements, since
not all vertices with four lines are allowed. Therefore,
\begin{eqnarray}
T_2 &=& 2\sum_{ij}z_{ij}^2
\times\epsfig{file=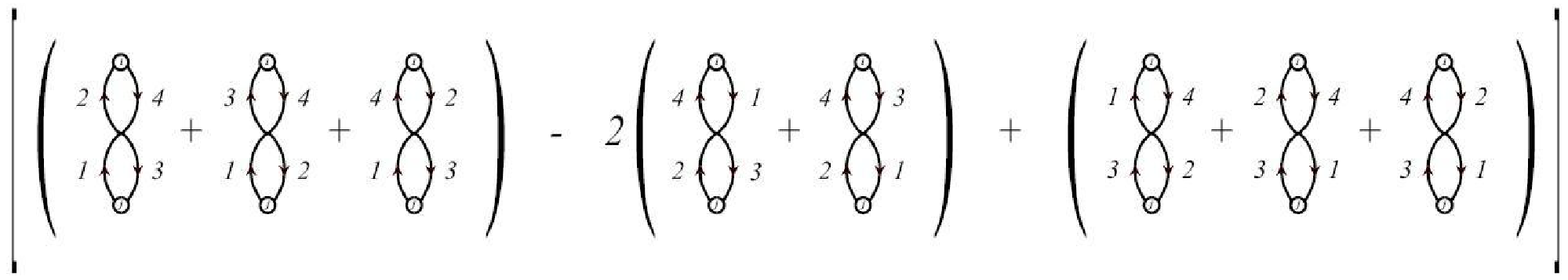,width=50ex}
\label{trace-term-d-expanded}
\end{eqnarray}
The product of ordered cumulants is the same for each diagram in
\eq{trace-term-d-expanded}. It is $\lbb +- \rbb^2 \lbb ++-- \rbb =
(\frac{1}{2})^2(-\frac{1}{2}) = -\frac{1}{8}$. Multiplying by
$\frac{1}{4}$ for the two circles, we obtain
\begin{eqnarray}
T_2 &=& 2\lp -\frac{1}{8} \rp \lp \frac{1}{4} \rp
\sum_{ijk}z_{ij}^2B_{ik}^2B_{jk}^2\times\lb 1(3) - 2(2) + 1(3) \rb
\nonumber \\ &=&  -\frac{1}{8}\sum_{ijk}z_{ij}^2B_{ik}^2B_{jk}^2.
\end{eqnarray}

The denominator of \eq{moment-binomial} is given by $\lbb +-
\rbb^2\sum_{ij}B_{ij}^2 = \frac{1}{4}\sum_{ij}B_{ij}^2$. Inserting
these results into \eq{moment-binomial}, we obtain
\begin{equation}
M_\cH^{(2)} =
-\frac{k^2}{4}\frac{\sum_{ij}z_{ik}^2B_{ij}^2B_{jk}^2}{\sum_{i}B_{ik}^2},
\label{MH2}
\end{equation}
where we have used translational invariance to drop one of the
summations.

The types of diagrams arising in the calculation of the fourth
moment are shown in Fig. \ref{MH4-diagrams}. To save space, the
distinct topologies are pictured without circles. The entire set
of diagrams at fourth order is obtained by placing two circles at
the vertices of the diagrams in Fig. \ref{MH4-diagrams} in all
possible ways. The result is straightforward to calculate, and is
\begin{eqnarray}
M_\cH^{(4)} &=& k^2\sum_{ij}z_{ik}^2B_{ik}^2B_{jk}^2 -2k^2
\frac{\sum_{ij} z_{ik}^2 \lp B_{ik}^2B_{jk}^4 + B_{ik}^4 B_{jk}^2
\rp}{\sum_{i} B_{ik}^2} -\frac{9}{4} k^2
\frac{\sum_{ij}z_{ik}^2B_{ik}^2B_{jk}^2B_{ij}^2}{\sum_i B_{ik}^2}
\nonumber \\
&& -\frac{k^2}{4}\frac{\sum_{ijl} z_{ik}^2 \lp
6B_{ik}^2B_{jk}B_{kl}B_{ij}B_{il} -
18B_{ik}B_{jk}^2B_{kl}B_{ij}B_{jl} +
11B_{jk}B_{kl}B_{ij}B_{il}B_{jl}^2\rp}{\sum_i B_{ik}^2}. \nonumber
\\ \label{MH4}
\end{eqnarray}
The sums over the index $k$ are left off, as usual. The first term
in \eq{MH4} comes from diagrams $a)$ and $b)$. Diagram $c)$ is of
the same order of magnitude, and gives the second term in this
equation. Diagrams $d)$ and $e)$ give rise to the third term, and
are an order of magnitude smaller for short-ranged or dipolar
coupling. Diagrams $g)$, $h)$, and $i)$ give the last term in
\eq{MH4} and are another order of magnitude smaller. The general
guidelines are that those diagrams with the greatest number of
lines per pair of vertices are the largest. The ones with several
pairs of vertices joined by only a single line, such as diagrams
$g)$, $h)$, and $i)$, are the smallest. There are exceptions to
these guidelines (For example, diagram $f)$ vanishes, for the same
reason as does the corresponding diagram at second order.), so
care must be taken in their application. As for the second moment,
the diagrams $l)$ vanish, as do diagrams $j)$ and $k)$.

\section{Numerical results for dipolar-coupled XY model}
\label{moments-numerics}

The results of numerical evaluation of the moments calculated in
the last section for $B_{ij} = B^{dip}_{ij}$ (see \eq{Bdipij}),
are given in Table\ \ref{moments-results}. This corresponds to
dipolar coupling. We have used values of the gyromagnetic ratio
and lattice spacing for the fluorines in calcium fluoride of
$\gamma = 2.51\times 10^{4} \; {\rm rad \; s^{-1}\; Oe^{-1}}$ and
$a = 2.73\times 10^{-8}\,{\rm cm}$. Because lattice sums can be
evaluated numerically only for finite lattice sizes, we used
finite size scaling to extract the infinite lattice limit. The
approach to the infinite lattice value is expected to follow a
power law. For example, if we approximate the sums by integrals in
Eq.\ (\ref{MZ2}),
\begin{eqnarray}
M_{\cM}^{(2)} &\approx& -k^2\int_{a \leq r \leq L}d^3\br \,
B^{dip}(\br)^2z^2 \nonumber \\ &\sim& {\rm const} \times \int_a^L
r^2dr \lp \frac{1}{r^3}\rp^2r^2 \nonumber \\ &=& {\rm const}
\times \lp \frac{1}{a} - \frac{1}{L} \rp. \label{scaling}
\end{eqnarray}
We performed a least squares fit to a power law of the quantities
in Eqs.\ (\ref{MZ2}), (\ref{MZ4}), (\ref{MH2}), and (\ref{MH4}) as
a function of lattice size, for both the [001] and [111]
orientations of the crystal with respect to the external field. We
found it sufficient to vary the lattice size between 1 and 81
lattice sites on an edge, in increments of 2 lattice sites. This
gave agreement with \eq{scaling} to better than one percent. The
numbers in Table\ \ref{moments-results} are the infinite lattice
values extracted from the scaling analysis.

\begin{table}
\begin{center}
\caption{Summary of the results for the dipolar coupled XY model
obtained from the moment method, with recent experimental values
for comparison.} \vspace{2ex}
\begin{tabular} {||c|c|c|c||} \hline Moments &[001]&[111]&  \\ \hline
\hspace*{0.1in} $M_{\cM}^{(2)}/k^2$ \hspace*{0.1in} ($ \times 10^{-7}$cm$^{2}$/s$^2$) & -5.59 & -2.21 & \\
\hspace*{0.1in} $M_{\cM}^{(4)}/k^2$ \hspace*{0.1in} ($ \times 10^{3}$cm$^{2}$/s$^4$) & 1.56 & 0.130 & \\
 \hline
\hspace*{0.1in} $M_{\cH}^{(2)}/k^2$ \hspace*{0.1in} ($ \times 10^{-7}$cm$^{2}$/s$^2$) & -2.80 & -1.08 & \\
\hspace*{0.1in} $M_{\cH}^{(4)}/k^2$ \hspace*{0.1in} (cm$^{2}$/s$^4$) & 76.2 & 28.4 & \\
 \hline Results for Gaussian cutoff & [001] &[111]&$D_{001}/D_{111}$ \\
\hline \hspace*{0.1in} $D_{\cM}$ \hspace*{0.1in}  ($ \times
10^{-12}$cm$^{2}$/s) & 13.3 & 11.4 & 1.17
\\ \hspace*{0.1in} $D_{\cH}$ \hspace*{0.1in}  ($ \times
10^{-12}$cm$^{2}$/s) & 21.2 & 8.4 & 2.5
\\ Ratio $D_{\cH}/D_{\cM}$& 1.59 & 0.74 & \\
 \hline \hspace*{0.1in} $T_{\cM}$ \hspace*{0.1in}  ($ \times
10^{-6}$ s) & 13.4 & 35.8 & \\ \hspace*{0.1in} $T_{\cH}$
\hspace*{0.1in}  ($ \times 10^{-6}$ s) & 42.8 & 43.8 &
\\
 \hline  Experiments (Refs.\ \cite{Zhang,Boutis}) & [001] &[111]&$D_{001}/D_{111}$ \\
\hline \hspace*{0.1in} $D_{\cM}$ \hspace*{0.1in} Ref. \cite{Zhang} $(\times 10^{-12}$ cm$^2$/s) & $7.1 \pm 0.5$ & $5.3 \pm 0.3$ & $1.34 \pm 0.12$\\
\hspace*{0.1in} $D_{\cH}$ \hspace*{0.1in} Ref. \cite{Boutis}
$(\times 10^{-12}$
cm$^2$/s) & $29 \pm 3$ & $33 \pm 4$ & $0.88 \pm 0.14$\\
Ratio $D_{\cH}/D_{\cM}$& $4.1 \pm 0.7$ & $6.2 \pm 1.1$ & \\
 \hline
\end{tabular}
\label{moments-results}
\end{center}
\end{table}

Besides the moments, Table\ \ref{moments-results} gives the values
for the diffusion coefficients for both Gaussian and step-function
cutoff (see \eq{alphas}), as well as their ratio. We find fair
agreement with experiments on calcium fluoride for the magnitudes
of both diffusion coefficients. For magnetization, our value is
slightly high, while for spin-spin energy it is slightly low. The
ratio $D_{\cH}/D_{\cM}$ that we calculate is about 1.6 for the
[001] direction, while in these experiments it is between 4 and 6.
Given the phenomenological nature of the theory we feel this to be
adequate agreement. For the [111] direction, the results are quite
different, giving a ratio of diffusion coefficients that is less
than one. We cannot account for this difference but conjecture
that it may be the result of neglecting the Ising, or $I^zI^z$,
term from the calculation.

As an additional check for consistency of this theory we have
calculated the value of the short time cutoff, $T_S$, using its
relation\cite{RY} to the moments of the appropriate cutoff
function in \eq{alphas}. As Table\ \ref{moments-results} shows,
$T_S$ was found to be on the order of 10 - 100 $\rm{ \mu s}$ for
the different cutoff functions and crystal orientations that we
considered. This is consistent with the assumption that $T_S$ is
related to the spin-spin correlation time given by the free
induction decay. The timescale associated with this decay in
calcium fluoride is approximately 20 ${\rm \mu s}$ with the
external field in the [001] direction and approximately 50 ${\rm
\mu s}$ with the external field in the [111] direction.\cite{LN}

\section{Conclusion}

We have shown how the diagrammatic technique for calculating
equilibrium correlation functions in spin systems may be adapted
to the evaluation of multi-spin dynamical correlation functions.
We used this technique to obtain exactly the first two
non-vanishing moments of the magnetization and spin-spin energy
autocorrelation functions of the XY model at infinite temperature
and long wavelength. The results were used to estimate the
magnetization and spin-spin energy diffusion coefficients in the
case of dipolar coupling, using a phenomenological moment method.
We found qualitative agreement with experiments on calcium
fluoride for both diffusion coefficients. The ratio of the
diffusion coefficient for spin-spin energy to that for
magnetization was found to be greater than one for the [001]
orientation of the external field with respect to the crystal
axes. However, this is not large enough to accurately account for
the observations. The orientation dependence of the diffusion
coefficients was also in qualitative agreement for magnetization,
but not for spin-spin energy. The lack of any experimentally
observed orientation dependence for spin-spin energy diffusion
leads us to conjecture that some additional, possibly
$k$-dependent, decay processes may have been at play in the
experiment, increasing the observed decay rates. Some artifacts of
the coherent time evolution of the spin system could also have
been involved, and would not be accounted for in the
phenomenological model of irreversible decay that was used here.
Finally, it is possible that the approximation of dropping the
Ising ($I^zI^z$) term was too drastic. A tractable calculation
including this term should be possible along the lines presented
here. Although we focused here on the spin diffusion problem, the
generality of the technique should allow for wider applicability.

\section{Acknowledgments} I would like to thank David Cory and
Chandrasekhar Ramanathan for introducing me to the problem of spin
diffusion and for their guidance in the early stages of this work.
I would also like to acknowledge support from the NSF through a
research assistantship at MIT and from a Feinberg Fellowship at
the Weizmann Institute.

\appendix

\section{Ordered Cumulants}
\label{cumulant-app}

Ordered Cumulants are used to simplify the evaluation of averages
of spin operators summed over the lattice. The non-zero elements
of a spin operator average contain, in general, several operators
with the same lattice site index. Since operators with different
indices commute, we may rearrange them so that all operators with
the same index are next to each other, and then factor the average
into averages over operators at different lattice sites, since
traces at different lattice sites are independent. For example,
$\langle I_k^+ I_i^z I_k^- I_i^z \rangle = \langle I^+_kI^-_k
\rangle \langle I^z_iI^z_i \rangle$ if $i\neq k$. Since we only
consider a Hamiltonian that is invariant under lattice
translations, the averages in the last expression are independent
of index. A general spin operator average may be calculated by
grouping the operators by index in this fashion, in all possible
ways, taking care to avoid over-counting by not including
identical groupings more than once. For example,
$\sum_{ik}A_{ik}\langle I_i^+I_k^- \rangle = \sum_{ik}
A_{ik}\lb\delta_{ik} \langle I_i^+I_i^- \rangle + (1-\delta_{ik})
\langle I_i^+ \rangle \langle I_k^- \rangle\rb = \sum_i A_{ii}\lp
\langle I_i^+I_i^- \rangle - \langle I_i^+ \rangle \langle I_i^-
\rangle \rp + \sum_{ik}A_{ik}\langle I_i^+ \rangle \langle I_k^-
\rangle$. Defining the ordered cumulants, $\lbb +- \rbb \equiv
\langle I_i^+I_i^- \rangle - \langle I_i^+ \rangle \langle I_i^-
\rangle$, $\lbb + \rbb \equiv \langle I_i^+ \rangle$, and $\lbb -
\rbb \equiv \langle I_i^- \rangle$, for an arbitrary index $i$, we
obtain $\sum_{ik}A_{ik}\langle I_i^+I_k^- \rangle = \lbb +- \rbb
\sum_i A_{ii} + \lbb + \rbb\lbb - \rbb\sum_{ik}A_{ik}$.

Generalizing the above example, we define ordered cumulants, also
known as semi-invariants,\cite{AG,Brout1,Brout2,Englert,SHEB}
iteratively in terms of their factorization in cumulants of lower
degree. Thus,
\begin{eqnarray}
\lbb + \rbb &=& \langle I^+ \rangle, \nonumber \\
\lbb - \rbb &=& \langle I^- \rangle, \nonumber \\
\lbb z \rbb &=& \langle I^z \rangle, \nonumber \\
\lbb I^zI^+ \rbb &=& \langle I^z I^+ \rangle - \lbb I^z \rbb
\lbb I^+ \rbb, \nonumber \\
\lbb I^zI^- \rbb &=& \langle I^z I^- \rangle - \lbb I^z \rbb
\lbb I^- \rbb, \nonumber \\
\lbb I^+I^- \rbb &=& \langle I^+ I^- \rangle - \lbb I^+ \rbb
\lbb I^- \rbb, \nonumber \\
\lbb I^zI^+I^- \rbb &=& \langle I^zI^+I^- \rangle - \lbb
I^zI^+\rbb\lbb I^- \rbb - \lbb I^zI^- \rbb \lbb I^+ \rbb \nonumber
\\&& - \lbb I^+I^-\rbb\lbb I^z\rbb -\lbb I^z\rbb\lbb I^+\rbb \lbb
I^-\rbb,
\end{eqnarray}
and so on. Ordered cumulants are related to spin operator averages
in an analogous way to the relation of cumulants and averages in
probability theory. The main difference is that the order of the
spin operators within the cumulant is important due to their
non-trivial commutation relations. Using ordered cumulants, it is
possible to calculate operator averages without restricting the
summation indices, as shown in the preceding paragraph.

\begin{table}
\caption{Ordered Cumulants for spin 1/2 and $T=\infty$. We use the
shorthand notation $+$ for $I^+$, $-$ for $I^-$, and $z$ for
$I^z$. Cumulants that do not have the same number of raising and
lowering operators are zero, and are not included. We also include
only one of each set of cumulants that differ by a cyclic
permutation of its operators. As discussed in the text, these
cumulants are the same at $T=\infty$.} \label{cumulants}
\begin{center}
\begin{tabular}{llll}\hline
\\ $\lbb z \rbb$  & $= 0$ & \hspace*{15ex} $\lbb +-+-z \rbb$  & $= 0$\\\\
$\lbb zz \rbb$  & $= \frac{1}{4}$ & \hspace*{15ex} $\lbb -++-z
\rbb$  & $= 0$
\\\\ $\lbb +- \rbb $ & $= \frac{1}{2}$ & \hspace*{15ex} $\lbb +--+z \rbb$  & $= 0$\\\\
$\lbb +-z \rbb$  & $= \frac{1}{4}$ & \hspace*{15ex} $\lbb ++--z \rbb$  & $= -\frac{1}{2}$ \\\\
$\lbb -+z \rbb$  & $= -\frac{1}{4}$ & \hspace*{15ex} $\lbb --++z
\rbb$  & $=
\frac{1}{2}$ \\\\
$\lbb zzz \rbb$  & $= 0$ & \hspace*{15ex} $\lbb -+-+z \rbb$  & $= 0$ \\\\
$\lbb zzzz \rbb$  & $= -\frac{1}{8}$ & \hspace*{15ex} $\lbb +++--- \rbb$  & $= \frac{3}{2}$\\\\
$\lbb +-zz \rbb$  & $= 0$ & \hspace*{15ex} $\lbb ++-+-- \rbb$  & $= \frac{1}{2}$\\\\
$\lbb +z-z \rbb$  & $= -\frac{1}{4}$ & \hspace*{15ex} $\lbb ++--+-
\rbb$  & $=
\frac{1}{2}$\\\\
$\lbb +-+- \rbb$  & $= \frac{1}{2}$ & \hspace*{15ex} $\lbb +-+-+- \rbb $ & $ = \frac{1}{2}$\\\\
$\lbb ++-- \rbb$  & $= -\frac{1}{2}$ & & \\\\
\hline
\end{tabular}
\end{center}
\end{table}

A list of ordered cumulants up to degree 5 for spin $\frac{1}{2}$
and $T=\infty$ is given in Table \ref{cumulants}. We omit
cumulants that differ only by a cyclic permutation of their
operators. In the limit of infinite temperature in which we are
interested, the density matrix is proportional to unity, and these
cumulants are the same by the properties of the trace. We note
that this cyclic invariance is not a general property at finite
temperature. Besides cyclic permutations, cumulants differing by
any other rearrangements in the order of the operators generally
have different values even at infinite temperature. Finally,
cumulants with unequal numbers of raising and lowering operators
are zero and are not included in the table.

\section{Cancellation of disconnected diagrams}
\label{appb}

We present a combinatorial proof of the cancellation of
disconnected diagrams for the time-independent Hamiltonian,
\eq{hamiltonian}. To account for time dependence or an $I^zI^z$
term, it is possible to proceed by the standard method via the
interaction picture and $S$-matrix expansion.\cite{Peskin}
However, the inapplicability of Wick's theorem for spin operators
prohibits the factorization of time-ordered products into
contractions, and we must eventually use the same type of counting
argument presented here.

\eq{moment-binomial} contains the term
\begin{equation}
\sum_{m=0}^n \lp \begin{array}{c} n \\
m \end{array}\rp (-1)^m \langle \cH^{n-m}S_j \cH^m S_i \rangle.
\label{corrA-binomial}
\end{equation}
There are two types of disconnected diagrams which contribute to
this term, those in which both $S_j$ and $S_i$ appear in the same
cumulant, and those in which they belong to different cumulants.
The latter type of disconnected diagram is always zero, because
the cyclic permutation symmetry of the trace allows us to factor
all the cumulants to the left of the summation over $m$ in
\eq{corrA-binomial}.

The case where both $S_j$ and $S_i$ appear in the same cumulant is
slightly more involved. Consider the subset of diagrams for which
$l < n$ interaction lines form the part which is not connected to
that containing $S_j$ and $S_i$. The $l$ interaction lines can
correspond to any of the $n$ $\cH$'s appearing in
\eq{corrA-binomial}, whose average may be factored outside the
summation over $m$. Depending on which ones we choose to factor
out, there will be a different number of $\cH$'s to the right of
the operator $S_j$. If we choose to leave $k < n$ $\cH$'s to the
right of
$S_j$, we can do this in $\lp \begin{array}{c} m \\
m - k \end{array}\rp \lp \begin{array}{c} n - m \\
l - (m-k) \end{array}\rp$ ways. The sum over $m$ in
\eq{corrA-binomial} for the set of diagrams with $l$ $\cH$'s
factored out is therefore equal to
\begin{eqnarray}
\sum_{m=0}^n && \lp \begin{array}{c} n \\
m \end{array}\rp (-1)^m \langle \cH^{n-m}S_j \cH^m S_i \rangle
\nonumber \\ &=& \langle \cH^l \rangle \sum_{k=0}^{n-l}
\langle \cH^{n-l-k}S_j\cH^kA_2\rangle \nonumber \\ &&\times\sum_{m=k}^{l+k} (-1)^m \lp \begin{array}{c} n \\
m \end{array}\rp \lp \begin{array}{c} m \\
m - k \end{array}\rp \lp \begin{array}{c} n - m \\
l - (m-k) \end{array}\rp. \nonumber \\ \label{appa-bigeq}
\end{eqnarray}
The product of binomial coefficients in this equation is
\begin{eqnarray}
\lp \begin{array}{c} n \\
m \end{array}\rp && \lp \begin{array}{c} m \\
m - k \end{array}\rp \lp \begin{array}{c} n - m \\
l - (m-k) \end{array}\rp \nonumber \\ &=&
\frac{n!m!(n-m)!}{(n-m)!m!(m-k)!k!(l+k-m)!(n-l-k)!}
\nonumber \\
&=& \frac{n!}{(m-k)!k!(l+k-m)!(n-l-k)!}.
\end{eqnarray}
The only factors that depend on $m$ are $\frac{1}{(m-k)!(l+k-m)!}
= \frac{1}{l!}\lp \begin{array}{c} l \\
m-k \end{array}\rp$. The sum over $m$ in \eq{appa-bigeq} is
therefore $\sum_{m=k}^{l+k}(-1)^m \lp \begin{array}{c} l \\
m-k \end{array}\rp = 0$. This proves the vanishing of disconnected
diagrams.


\begin{thebibliography}{00}
\bibitem{Lorentzian} The truncated
Lorentzian leading to \eq{D} has also been used in the theory of
resonance lineshapes in solids. The underlying physical assumption
is the same in both cases, namely that correlations decay
exponentially at long times and quadratically at short times. See,
for instance Ref.\ \cite{Abragam}, Ch. 3.
\end{thebibliography}
\end{document}